\documentclass[prd,twocolumn,superscriptaddress,altaffilletter,nofootinbib]{revtex4}

\usepackage{cancel}
\usepackage{graphicx}
\usepackage{amsmath}
\usepackage{amssymb}
\usepackage{graphicx,epsfig}
\newtheorem{theorem}{Theorem}
\newtheorem{corollary}{Corollary}[theorem]

\newcommand{\be}{\begin{equation}}
\newcommand{\ee}{\end{equation}}
\newcommand{\bea}{\begin{eqnarray}}
\newcommand{\eea}{\end{eqnarray}}
\newcommand{\der}{\partial}
\newcommand{\vphi}{\varphi}
\newcommand{\bet}{\begin{theorem}}
\newcommand{\eet}{\end{theorem}}
\newcommand{\bec}{\begin{corollary}}
\newcommand{\eec}{\end{corollary}}

\begin{document}



\title{Comment on ``Dark matter as a Weyl geometric effect''}



\author{Israel Quiros}\email{iquiros@fisica.ugto.mx}\affiliation{Dpto. Ingenier\'ia Civil, Divisi\'on de Ingenier\'ia, Universidad de Guanajuato, Gto., M\'exico.}



\begin{abstract}In this note we comment on a recent attempt by P. Burikham, T. Harko, K. Pimsamarn and S. Shahidi [Phys. Rev. D {\bf 107}, 064008 (2023)] to explain the galactic rotation curves as the result of the motion of time-like test particles in the Weyl geometric theory of gravity. We show that the static, spherically symmetric solution found by the authors, which could be the basis of an alternative explanation of the galactic rotation curves, is wrong.\end{abstract}



\maketitle




\section{Wrong vacuum solution}\label{sect-sol}


The equations of motion (EOM) of WGG theory are the following: i) The Einstein's EOM,

\bea &&{\cal E}_{\mu\nu}:=G_{\mu\nu}-\frac{1}{\phi^2}\left(\nabla_\mu\nabla_\nu-g_{\mu\nu}\nabla^2\right)\phi^2\nonumber\\
&&\;\;\;\;\;\;\;-\frac{3}{2}\left(w_\mu w_\nu-\frac{1}{2}g_{\mu\nu}w_\lambda w^\lambda\right)\nonumber\\
&&\;\;\;\;\;\;\;+\frac{3}{2}\left(\der_{\mu}\phi^2w_{\nu}+\der_{\nu}\phi^2w_{\mu}-g_{\mu\nu}\der_\lambda\phi^2w^\lambda\right)\nonumber\\
&&\;\;\;\;\;\;\;-\frac{\beta^2}{\phi^2}\left(F^{\;\;\lambda}_\mu F_{\nu\lambda}-\frac{1}{4}g_{\mu\nu}F^2_{\lambda\sigma}\right)=-\Lambda\phi^2g_{\mu\nu},\label{einst-eom}\eea ii) The ``Klein-Gordon'' equation for the scalar field,

\bea R-\frac{3}{2}w_\mu w^\mu-3\nabla_\mu w^\mu=4\Lambda\phi^2,\label{phi-eom}\eea and iii) The inhomogeneous Proca equation for the nonmetricity vector field,

\bea \nabla^\nu F_{\mu\nu}=-\frac{3}{2\beta^2}\phi^2\left(w_\nu-\frac{\der_\nu\phi^2}{\phi^2}\right).\label{max-eom}\eea In the above equations $\Lambda$ and $\beta^2$ are free coupling parameters. Equations \eqref{einst-eom}, \eqref{phi-eom} and \eqref{max-eom}, are invariant under the Weyl gauge transformations: $g_{\mu\nu}\rightarrow\Omega^2g_{\mu\nu},$ $w_\mu\rightarrow w_\mu-2\der_\mu\ln\Omega,$ where the positive function $\Omega$ is the conformal factor. In what follows, for brevity, we shall call these transformations just as ``gauge transformations.'' If set the constant $\Lambda=1/4$ in equations \eqref{einst-eom} and \eqref{phi-eom}, there is a full correspondence between the above vacuum equations and the corresponding equations in \cite{harko-prd-2023}. However, this choice makes difficult to take the limit $\Lambda\rightarrow 0$, so that, unless otherwise stated, in this comment we leave $\Lambda$ to be a free constant.


In \cite{harko-prd-2023} the authors look for static, spherically symmetric solutions to the vacuum equations \eqref{einst-eom}, \eqref{phi-eom} and \eqref{max-eom} (equations (24), (26) and (28) in \cite{harko-prd-2023}.) The following line element in spherical coordinates is considered:

\bea ds^2=-e^{\nu(r)}dt^2+e^{\lambda(r)}dr^2+r^2d\Omega^2,\label{line-element}\eea where $d\Omega^2\equiv d\theta^2+\sin^2\theta d\vphi^2$. The analysis in \cite{harko-prd-2023} is restricted to the following choice of the nonmetricity vector (Eq. (31) in \cite{harko-prd-2023}):

\bea w_\mu=(0,w_r,0,0).\label{nm-vect}\eea This choice entails that $F_{\mu\nu}=0$. This means, in turn, that the nonmetricity vector must be the gradient of some scalar: $w_\mu=\nabla_\mu f$. As a matter of fact, if substitute $F_{\mu\nu}=0$ into the Proca equation \eqref{max-eom}, one gets that $w_\mu=\der_\mu\phi^2/\phi^2$. Hence, the nonmetricity law of Weyl space $\tilde W_4$ transforms into the nonmetricity law defining WIG space $\tilde W^\text{int}_4$: $\hat\nabla_\alpha g_{\mu\nu}=-(\der_\alpha\phi^2/\phi^2)g_{\mu\nu}.$ Besides, vanishing of the nonmetricity field strength: $F_{\mu\nu}=0$, transforms equation \eqref{einst-eom} into the following simplified EOM:

\bea &&{\cal R}_{\mu\nu}:=R_{\mu\nu}+\frac{6}{\phi^2}\der_\mu\phi\der_\nu\phi\nonumber\\
&&\;\;\;\;\;\;\;\;-\frac{1}{\phi^2}\left(\nabla_\mu\nabla_\nu+\frac{1}{2}g_{\mu\nu}\nabla^2\right)\phi^2=\Lambda\phi^2g_{\mu\nu},\label{vac-eq}\eea where we have conveniently substituted the curvature scalar $R$ into the Einstein's tensor $G_{\mu\nu}$, after taking the trace of \eqref{einst-eom}. Meanwhile the scalar field's EOM \eqref{phi-eom} simplifies to:\footnote{In order to exactly reproduce the equations in \cite{harko-prd-2023} we have to set $\Lambda=1/4$ in the RHS of equations \eqref{vac-eq} and \eqref{kg-eom}.} 

\bea R+6\frac{(\der\phi)^2}{\phi^2}-3\frac{\nabla^2\phi^2}{\phi^2}=4\Lambda\phi^2.\label{kg-eom}\eea The later is not an independent equation since it coincides with the trace of \eqref{vac-eq}. Hence, the scalar $\phi(x)$ is not a dynamical field: it is a free function which does not obey a specific EOM. This is a consequence of gauge symmetry, given that equations \eqref{vac-eq} and \eqref{kg-eom} are invariant under the following gauge transformations:

\bea g_{\mu\nu}\rightarrow\Omega^2g_{\mu\nu},\;\phi\rightarrow\Omega^{-1}\phi.\label{gauge-t'}\eea 


In \cite{harko-prd-2023} (see also \cite{harko-epjc-2022}) the following static, spherically symmetric solution of \eqref{vac-eq} was found:

\bea &&\phi(r)=\frac{\sqrt{C_1}}{r+C_2},\label{harko-sol'}\\
&&e^{\nu}=e^{-\lambda}=1-\delta+\frac{\delta(2-\delta)}{6m}r-\frac{2m}{r}+C_3r^2,\label{harko-sol}\eea where $\delta=3(1-C_3C^2_2)-C_1/4$, besides $\delta C_2=6m$. This means that the integration constants $C_1$, $C_2$ and $C_3$ are not independent of each other. The above equations coincide with equations (49) and (61) of Ref. \cite{harko-prd-2023}, respectively.

Despite that one of the declared aims of \cite{harko-prd-2023} is to search for a solution of WGG vacuum equations \eqref{einst-eom}, \eqref{phi-eom} and \eqref{max-eom}, what is actually done in that reference is to look for a static, spherically symmetric solution of the vacuum EOM \eqref{vac-eq}, which happens to be wrong.


\section{Shortcomings of the study in \cite{harko-prd-2023}}\label{sect-deficit}


There were shortcomings in the exact solutions study in Ref. \cite{harko-prd-2023}, which led to the wrong result. 

First, it was not noticed in the mentioned reference that, due to gauge symmetry, the scalar field is a free function. The ``Klein-Gordon'' equation for the scalar field (Eq. (27) in \cite{harko-prd-2023}):

\bea \nabla_\mu(\Phi w^\mu)=\nabla^2\Phi,\label{harko-kg}\eea where $\Phi=\phi^2$, is not an independent equation. Actually, due to the form assumed by the nonmetricity vector: $w_\mu=(0,w_r,0,0)$, the nonmetricity field tensor vanishes: $F^{\mu\nu}=0$. If substitute this last condition in \eqref{max-eom}, we get that $\Phi w^\mu=\nabla^\mu\Phi$, which when substituted in the left-hand side (LHS) of \eqref{harko-kg}, leads to the identity: $\nabla^2\Phi=\nabla^2\Phi$. Means that Eq. \eqref{harko-kg} can not be the scalar field's EOM. In consequence, the gauge condition $\nabla_\mu w^\mu=0$ (Eq. (36) of Ref. \cite{harko-prd-2023},) amounts to an additional equation:

\bea \frac{\Phi''}{\Phi}+\left(\nu'-\lambda'+\frac{4}{r}\right)\frac{\Phi'}{2\Phi}-\left(\frac{\Phi'}{\Phi}\right)^2=0,\nonumber\eea which forces a specific solution for the scalar field, i. e., it amounts to fixing a gauge. This is why in \cite{harko-prd-2023} a solution for $\Phi=\phi^2(x)$ is found (Eq. \eqref{harko-sol'} above.)

Second, the assumption (Eq. (47) in \cite{harko-prd-2023}):

\bea \nu(r)+\lambda(r)=0,\label{fatal-c}\eea is incorrect in general. At least it is not compatible with the equations of motion \eqref{vac-eq}. Actually, if substitute this condition in Eq. (42) of Ref. \cite{harko-prd-2023}, the latter can be written in the following way:

\bea \frac{\Phi'}{\Phi}=-\frac{C_1e^{-\nu}}{r^2},\label{42}\eea where $C_1$ is an integration constant. If further substitute in \eqref{42} the scalar field's solution \eqref{harko-sol'}: $\Phi=C_1/(r+C_2)^2$, we get that:

\bea e^\nu=\frac{C_1}{2r}\left(1+\frac{C_2}{r}\right),\label{wrong}\eea which does not coincide with the solution \eqref{harko-sol}, for the metric coefficients. Eq. \eqref{harko-sol} is obtained after integrating Eq. (54) of Ref. \cite{harko-prd-2023}, which can be written as (here we introduce the variable $u=re^\lambda$):

\bea \frac{C_2}{C_2+r}u'-\frac{3C_2}{(C_2+r)^2}u+\frac{C_1 r^2}{4(C_2+r)^2}-1=0.\label{harko-diff-eq}\eea This equation is obtained by substituting $\Phi$ from \eqref{harko-sol'} into Eq. (53) in \cite{harko-prd-2023}. In other words: The assumption \eqref{fatal-c}, is incompatible with equations \eqref{harko-sol'} and \eqref{harko-diff-eq}. 

This inconsistency is due to the fact that in \cite{harko-prd-2023} it were made more assumptions than required by the symmetries of the solution: There are three vacuum equations and three unknown functions: $\nu$, $\lambda$ and $\Phi$. However, only two of the vacuum equations are actually independent differential equations.\footnote{As an illustration, one may consider the GR-de Sitter vacuum equations, which are a particular case of \eqref{vac-eq} when $\phi=$const.$=1$. The resulting vacuum equations read: 1) $2\nu''+\nu'^2-\nu'\lambda'+4\nu'/r=-4\Lambda e^\lambda$, 2) $-2\nu''-\nu'^2+\nu'\lambda'+4\lambda'/r=4\Lambda e^\lambda$ and 3) $r(\lambda'-\nu')-2+2e^\lambda=2\Lambda e^\lambda$. Although it seems that there are three independent equations, the sum of the first two equations leads to the condition $4(\nu'+\lambda')/r=0$, which when substituted back into the vacuum equations, leads to two independent equations on the unknowns $\nu$ and $\lambda$, including the above condition itself. In other words: we can work with equations 1) and 2) exclusively, in order to solve for the unknown metric functions.\label{fnote}} Therefore, the gauge condition: $\nabla_\mu w^\mu=0$, amounts to an additional equation which allows to force a solution for the scalar field $\Phi$. There is no more freedom to make additional assumptions. Otherwise if make an additional assumption, as in Eq. \eqref{fatal-c}, the system of differential equations is overdetermined. Usually a condition like \eqref{fatal-c} results from the equations of motion and should not be imposed a priori (see footnote \ref{fnote}.)



\section{Discussion}\label{sect-discu}


The most important aim of \cite{harko-prd-2023} is to find an alternative explanation of the galactic rotation curves in the WGG theory. Galaxies are modeled by a baryonic pressureless fluid. Means that gauge symmetry is broken. Yet, since the fluid is pressureless, it happens that the scalar field profile is the same as in the vacuum case. The only modification of the vacuum equations is the addition of a nonvanishing matter source term: $-\kappa\rho r^2$, in the RHS of Eq. \eqref{harko-diff-eq}, where $\rho$ is the baryonic matter density. The modification results in Eq. (97) of \cite{harko-prd-2023}: 

\bea \frac{C_2}{C_2+r}u'-\frac{3C_2}{(C_2+r)^2}u+\frac{C_1 r^2}{4(C_2+r)^2}-1=-\kappa\rho r^2.\label{97}\eea Notice that the LHS of this equation coincides with the LHS of \eqref{harko-diff-eq}, which is obtained with the help of equations \eqref{42} and \eqref{harko-sol'} (same as equations (42) and (49) of Ref. \cite{harko-prd-2023},) being still valid. When these equations are taken jointly with the condition \eqref{fatal-c}, one obtains Eq. \eqref{wrong}, which is not solution of \eqref{97}. The assumption \eqref{fatal-c} is not compatible with the equation of motion \eqref{97} either.

Given that the comparison of the theoretical WGG model with experimental data in sections IV and V of Ref. \cite{harko-prd-2023}, is based on equations which include the wrong assumption \eqref{fatal-c}, the results of the study should be taken with caution since these are, at least, misleading.


{\bf Acknowledgments.} The author acknowledges FORDECYT-PRONACES-CONACYT for support of the present research under grant CF-MG-2558591.





\end{document}